\documentstyle[12pt,twoside,fleqn,espcrc1,epsfig]{article}
\begin{document}
\newcommand{\ttbs}{\char'134}
\newcommand{\AmS}{{\protect\the\textfont2
A\kern-.1667em\lower.5ex\hbox{M}\kern-.125emS}}

\hyphenation{author another created financial paper re-commend-ed}

%
%
\title{Secondary Charmonium Production in Heavy Ion Collisions \\
at LHC Energy
}

\author{P. Braun-Munzinger\address{Gesellschaft f\"ur
        Schwerionenforschung, GSI
         \\
        Postfach 110552, D-64220 Darmstadt, Germany}%
       ~and
        K. Redlich$^{a,}$\address{Institute  of Theoretical Physics,
        University of Wroc\l aw, \\
        PL-50204 Wroc\l aw, Poland}}

\maketitle

\begin{abstract}
 We consider the production of charmonium by $D\bar D$ annihilation
 during the mixed and hadronic phase of a Pb-Pb collision
 at LHC energy.
 The calculations for secondary $J/\psi$ and $\psi^,$ production are
 performed within a kinetic model taking into account the space-time
 evolution of a longitudinally and transversely expanding medium. It
 is shown that the yield of secondary $J/\psi$ mesons depends strongly
 on the $J/\psi$ dissociation cross section with co-moving hadrons.
 Within the most likely scenario for the dissociation cross section it
 will be negligible. The secondary production of $\psi^,$ mesons,
 however, due to their large cross section above the threshold, can
 substantially exceed the primary yield.
\end{abstract}
\section{Secondary charmonium production}

The initial energy density in ultrarelativistic heavy ion collisions
at LHC energy exceeds by a few order of magnitudes the critical value
required for quark-gluon plasma formation. Thus, according to Matsui
and Satz \cite{1}, one expects the formation of charmonium bound
states to be severely supressed due to Debye screening. The initially
produced $c\bar c$ pairs in hard parton scattering, however, due to
charm conservation, will survive in the deconfined medium until the
system reaches the critical temperature where the charm quarks
hadronize forming predominatly $D$ and $\bar D$ mesons.  The
appreciable number of $c\bar c$ pairs and consequently $D$,$\bar D$
mesons expected in Pb-Pb collisions at LHC energy can lead to
additional production of charmonium bound states due to the reactions
such as, $D\bar {D^*} +D^*\bar D + D^*\bar {D^*}\to \psi +\pi$ and
$D^*\bar {D^*} +D\bar D \to \psi +\rho$ as first indicated in
\cite{2}.  In this work we present a quantitative description of
secondary $J/\psi$ and $\psi^,$ production due to the above processes
from the thermal hadronic medium created in Pb-Pb collisions at LHC
energy.

\section{Cross section and production rate}
\begin{figure}[htb]
\begin{center}
\epsfig{file=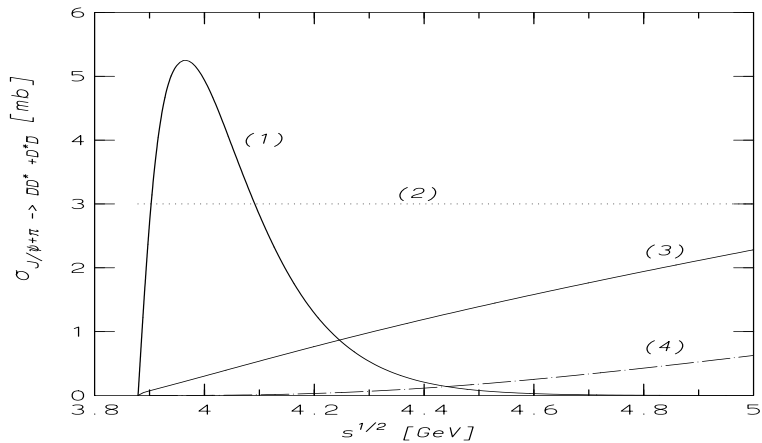, width=110mm,height=64mm}
\end{center}
\vspace*{-1.5 cm}
\caption{Energy dependence of the $J/\psi$ absorption
on pions predicted in the contex of four
different models (see text for description).
}
\label{fig:1}
\end{figure}
\begin{figure}[htb]
\vspace*{-1.0 cm}
\begin{center}
\epsfig{file=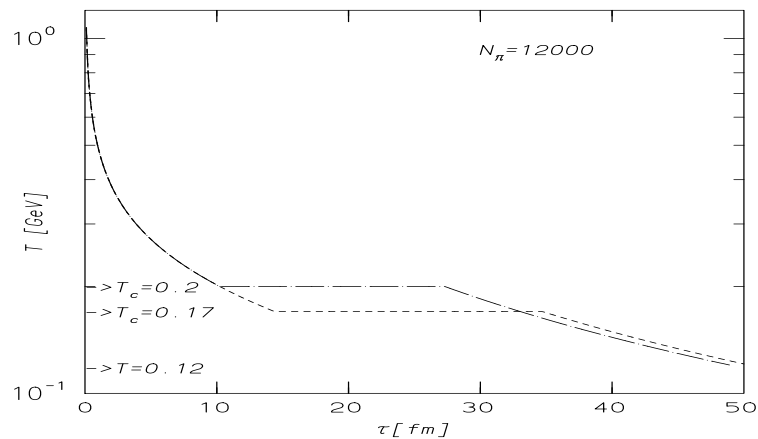, width=110mm,height=64mm}
\end{center}
\vspace*{-1.5 cm}
\caption{Model for space-time evolution of the hot medium created in Pb-Pb
collisions at LHC energy.}
\label{fig:2}
\vspace*{-0.6 cm}
\end{figure}
\begin{figure}[htb]
\begin{center}
\epsfig{file=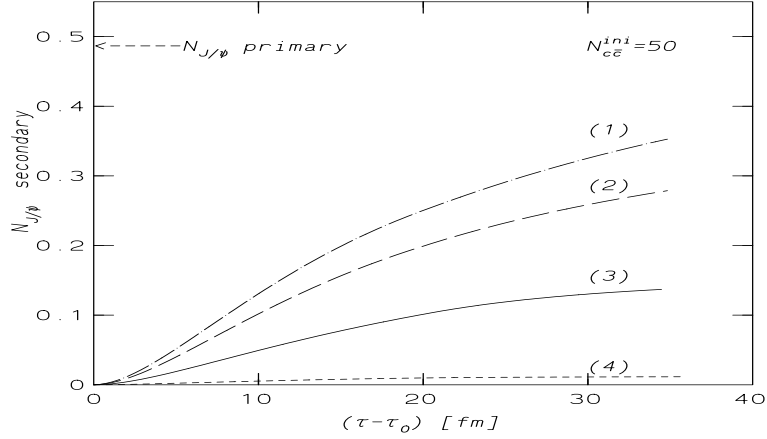, width=110mm,height=64mm}
\end{center}
\vspace*{-1.50 cm}
\caption{Time evolution of the abundance of
 secondarily produced $J/\psi$ mesons from Pb-Pb
collisions at LHC calculated with the  cross sections from fig.1.
}
\label{fig:3}
\end{figure}
\begin{figure}[htb]
\vspace*{-1.0 cm}
\begin{center}
\epsfig{file=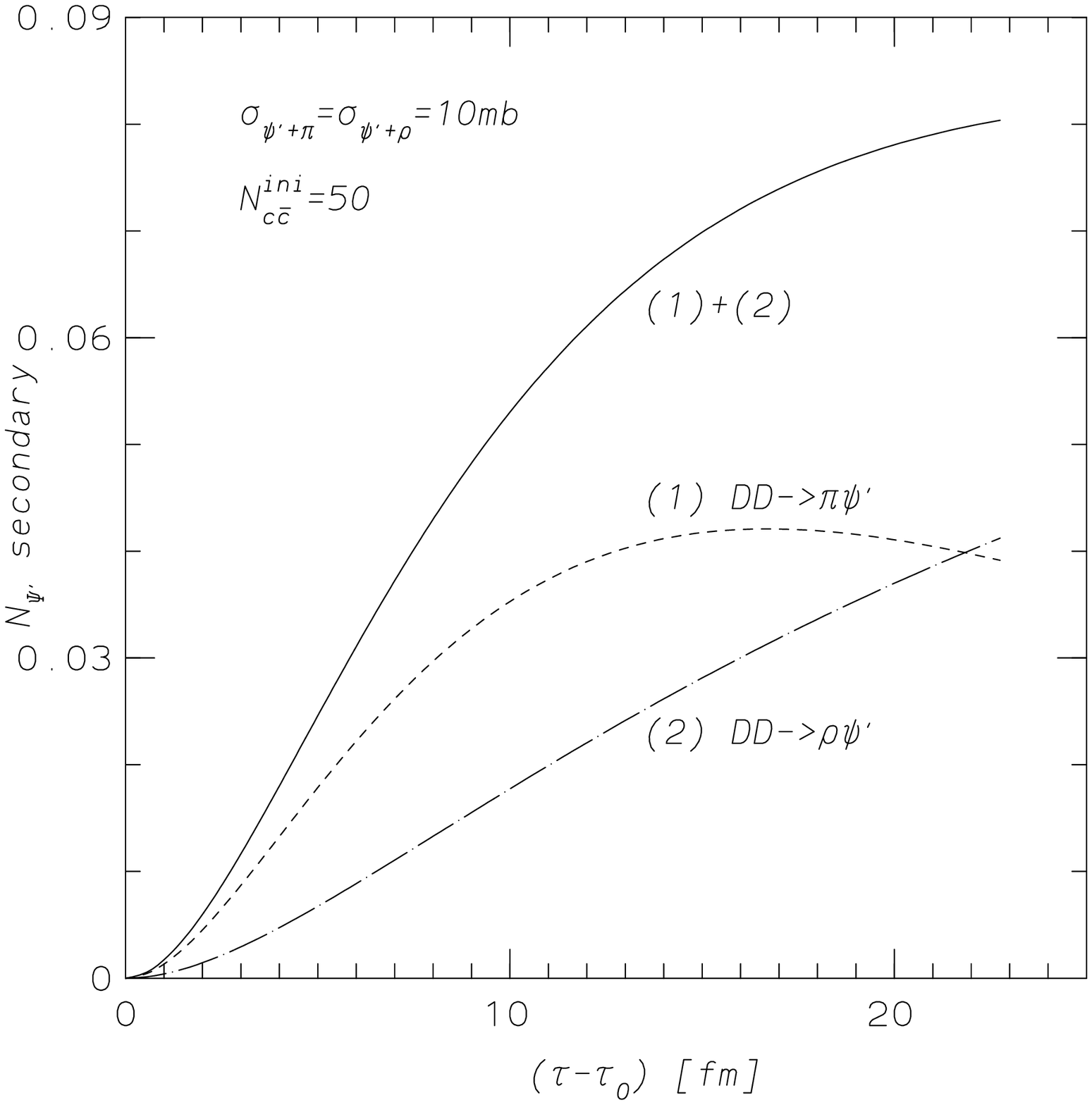, width=110mm,height=64mm}
\end{center}
\vspace*{-1.50 cm}
\caption{Time evolution of the abundance of $\psi^,$ from Pb-Pb
collisions at LHC calculated with constant absorption cross section,
$\sigma_{\psi^,\pi}\sim \sigma_{\Psi^,\rho}\sim 10$mb.}
\label{fig:4}
\vspace*{-0.6 cm}
\end{figure}

The charmonium production cross section $\sigma_{D\bar D\to \psi h}$
can be related to the hadronic absorption of charmonium $\sigma_{\psi
  h\to D\bar D}$, through detailed balance. The magnitude of
charmonium absorption cross section on hadrons is still, however,
theoretically not well under control. In fig.1 we show the predictions
for the dissociation cross section of $J/\psi$ on pions of: (1) the
quark exchange model \cite{3}, (2) the comover model with
$\sigma_{\psi\pi}\sim$3mb \cite{4}, (3) calculations using an
effective hadronic Lagrangian \cite{5} and (4) a short distance QCD
approach \cite{6}. The large theoretical uncertainties of the cross
section seen in fig.1 will naturally influence the yield of secondary
charmonium bound states.

In the thermal hadronic medium  the rate of charmonium production
 from $D\bar D$ annihilation is determined by
the thermal average of the cross section
and the densities of incoming and outgoing particles \cite{2}.
The solution of the rate equation requires additional assumptions
on    the space-time evolution of the hadronic medium  and the initial
number of $D$ and  $\bar D$ mesons  at the beginning of the mixed phase.

We have adopted a hydrodynamical model for expansion dynamics
assuming that at initial time $\tau_0\sim 0.1$fm the system
is created as an equilibrium quark-gluon plasma of temperature
$T_0\sim 1$GeV
and is then undergoing isentropic longitudinal expansion with a superimposed
transverse flow.
In fig.2 we show the time evolution of the temperature in our model
for two values of $T_c$ with 12000 pions (charged and neutral)per unit
of rapidity in the final state.  In the actual numerical calculation
we take $T_c\sim 0.17$GeV as value for the critical temperature,
following recent calculations within the lattice gauge theory approach
\cite{7}. The analysis of presently available data for the yield of
hadrons produced in heavy ion collisions suggests that the chemical
freezeout temperature at LHC should be in the range $0.16<T_f<0.17$
\cite{8,9}, i.e very close to $T_c$.

To get the initial number of
 $c\bar c$ pairs in Au-Au collisions at LHC we
scale the $p-p$ calculations
from PYTHIA with the total number of nucleon-nucleon collisions \cite{10}.
 Typical rapidity densities   of 50 $c\bar c$
 pairs were obtained, leading to about 50 $D$ and
 $\bar D$ mesons and 0.5 primary
 $J/\psi$ at midrapidity.
 
 In fig.3 we show the time evolution of the abundance of $J/\psi$
 mesons as obtained from the solution of the kinetic equations with
 four different values of the $J/\psi$ absorption cross section as
 described in fig.1. As is seen in fig.3 the yield is very sensitive
 to the size of the absorption cross section. For example, using the
 short distance QCD approach leads to negligible secondary production
 of $J/\psi$ mesons.  Similar analysis for $\psi^,$ is shown in fig.4.
 Here, due to the small binding energy of the $\psi^,$ meson and its
 correspondingly larger size, we have taken the cross section for
 $\psi^,$ absorption by pions and rho mesons to be energy independent
 and equal to its geometric value of 10 mb just at threshold
 \cite{11}. The results in fig.4 show that the secondary yield of
 $\psi^,$ can be large and reaches the value of almost 1/5 of the
 initial number of primary $J/\psi$. The yield is also seen to be very
 weakly dependent on the freezeout time.
 
 Thermal $c\bar c$ pairs produced during the evolution of quark-gluon
 plasma are, within our approach, found to increase the secondary
 charmonium yield shown in fig$^,$s.3,4 by 40$\%$ with an equilibrium
 initial conditions as used in fig.2 and by 20$\%$ if using the
 initial conditions from SSPC \cite{12} for a plasma out of chemical
 equilibrium.

 A detailed
presentation of the results can be found in \cite{13}.

\section{Conclusions}
We have shown that secondary charmonium production in heavy ion
collisions appears almost entirely during the mixed phase.  The yield
of secondarily produced $J/\psi$ mesons is very sensitive to the
hadronic absorption cross section. Within the context of the short
distance QCD approach this leads to negligible values for J/$\Psi$
regeneration.  The $\psi^,$ production, however, can be large and may
even exceed the initial yield from primary hard scattering.  It is
thus conceivable that at LHC energy the $\psi^,$ charmonium state can
be seen in the final state whereas $J/\psi$ production can be entirely
suppressed.

We  acknowledge stimulating discussions with H. Satz
and J. Stachel.

\end{document}